\documentclass[12pt]{iopart}
\usepackage{iopams,graphicx}

\usepackage{lmodern}
\usepackage[T1]{fontenc}
\usepackage{bbold}
\graphicspath{{Graphics/}}
\usepackage{xcolor}

\usepackage{soul}
\usepackage[mathscr]{euscript}
\usepackage{cite}
\usepackage[super]{nth}
\usepackage{float}  
\usepackage{comment}   
\usepackage[colorlinks, citecolor = blue, urlcolor = blue]{hyperref}

\begin{document}

\title[Reducing mean first passage times with intermittent confining potentials]{Reducing mean first passage times with intermittent confining potentials: a realization of resetting processes}

\author{Gabriel Mercado-V\'asquez$^1$, Denis Boyer$^1$,	and Satya N. Majumdar$^2$}

\address{$^1$ Instituto de F\'isica, Universidad Nacional Aut\'onoma de M\'exico, Mexico City 04510, Mexico}
\address{$^2$ LPTMS, CNRS, Univ. Paris-Sud, Universit\'e Paris-Saclay, 91405 Orsay, France}


  \begin{abstract}
      During a random search, resetting the searcher's position from time to time to the starting point often reduces the mean completion time of the process. Although many different resetting models have been studied over the past ten years, only a few can be physically implemented. Here we study theoretically a protocol that can be realised experimentally and which exhibits unusual optimization properties. A Brownian particle is subject to an arbitrary confining potential $v(x)$ which is switched on and off intermittently at fixed rates. Motion is constrained between an absorbing wall located at the origin and a reflective wall. When the walls are sufficiently far apart, the interplay between free diffusion during the "off" phases and attraction toward the potential minimum during the "on" phases gives rise to rich behaviours, not observed in ideal resetting models. For potentials of the form $v(x)=k|x-x_0|^n/n$, with $n>0$, the switch-on and switch-off rates that minimise the mean first passage time (MFPT) to the origin undergo a continuous phase transition as the potential stiffness $k$ is varied. When $k$ is above a critical value $k_c$, 
      potential intermittency enhances target encounter: 
      the minimal MFPT is lower than the Kramer's time and is attained for a non-vanishing pair of switching rates. We focus on the harmonic case $n=2$, extending previous results for the piecewise linear potential ($n=1$) in unbounded domains. We also study the non-equilibrium stationary states emerging in this process.
    \end{abstract}

\section{Introduction}

Resetting processes have attracted an increased attention over the last decade in the field of non-equilibrium statistical physics \cite{Evans_2020}. 
In the model originally introduced in \cite{EvansPRL2011}, a Brownian particle is instantaneously reset to its initial position at exponentially distributed time intervals. This rather simple modification of standard Brownian motion has important effects on several basic static and dynamical observables. Returning the particle from time to time to a specific position can accelerate the detection of an absorbing target by cutting off fruitless excursions where the particle diffuses far away from the target. For a Brownian particle in an unbounded domain, resetting makes the mean first passage time (MFPT) to an absorbing target finite, and this quantity can be further minimized by a suitable choice of the resetting rate $r$ \cite{Evans_2011,EvansPRL2011}. 
The effects of resetting have been explored on a wide range of random processes, such as diffusion in bounded domains\cite{Christou_2015,ray2021resetting,ahmad2022first}, anomalous diffusion\cite{Kusmierz2014PRL,KusmierzPRE2015,KusmierzPRE2019,MasoPRE2019}, searches with stochastically gated targets\cite{Bressloff_2020,Mercado_V_squez_2021}, or resetting under more general protocols \cite{Eule_2016,MonteroPRE2016,pal2016diffusion,nagar2016diffusion,boyer2017long, falcon2017localization,ChechkinPRL2018,evans2018effects}
(see \cite{Evans_2020,Shamik2022} for two reviews of this topic).

Resetting also gives rise generically to non-equilibrium stationary states (NESS)  \cite{EvansPRL2011,manrubia1999stochastic,mendez2016characterization}. For instance, the position density of the Brownian particle under stochastic resetting becomes effectively localized in space at large times and converges to a NESS that resembles a Boltzmann-Gibbs distribution with an effective confining potential given by $v_{eff}(X)=\sqrt{r/D}|X-X_0|$, where $D$ is the diffusion constant of the particle and $X_0$ the resetting position\cite{Evans_2011,EvansPRL2011}. However, the process itself markedly differs from the Langevin dynamics at equilibrium in this effective potential, and the first passage properties in both cases are consequently very different. On a semi-infinite line, the MFPT to an absorbing target of a Brownian motion under resetting is always less than the MFPT at equilibrium in the corresponding effective external potential \cite{Evans_2013,giuggioli2019comparison}. 

Given the advantages of search processes based on resetting over equilibrium dynamics, it is desirable to design protocols that allow Brownian particles to shorten their encounter times with a target and that are experimentally feasible. A few experiments on resetting processes have been carried out. They typically utilise microspheres manipulated with optical tweezers and different resetting protocols have been considered \cite{besga2020optimal,tal2020experimental,besga2021dynamical,Faisant_2021}. Resetting poses two important practical challenges in physical systems: (i) particles cannot be reset instantaneously and (ii) not exactly to the same position due to thermal or experimental fluctuations. In order to address these issues, different modifications to the original model have been proposed, such as non-instantaneous resetting \cite{maso2019transport,pal2019time,bodrova2020resetting,bodrova2020two,gupta2020stochastic,pal2019home} or resetting to a distribution of positions \cite{Evans_2011,besga2020optimal}.
Another convenient way of emulating resetting processes consists in using an external trapping potential that is alternatively switched on and off. During the "on" phases, the diffusing particle returns advectively towards the potential minimum (which plays a role similar to the resetting position in ideal resetting), before diffusing freely when the potential is no longer applied. The NESSs and the diffusion properties that emerge in stochastically switching harmonic potentials have been studied for Brownian particles along different schemes \cite{gupta2020stochastic,santra2021brownian}, as well as for L\'evy walks \cite{xu2022stochastic}. The work distribution \cite{gupta2022work} and entropy production have also been explored \cite{alston2022non}. Notwithstanding these advances, the properties of first passage times in fluctuating potentials remain little understood.

\begin{figure}[t]
\centering
			\includegraphics[width=\textwidth]{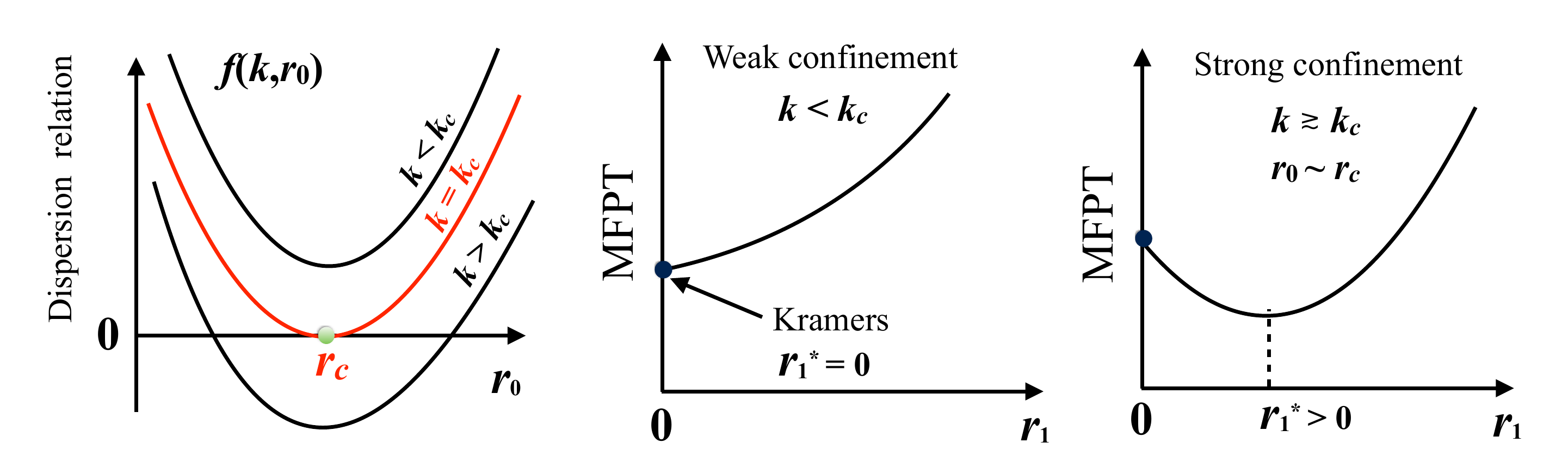}
            
			\caption{Brownian searches assisted by intermittent potentials. {\bf Left:} Leading order of the non-equilibrium contribution to the mean search time in Eq. (\ref{eq0}), when the potential stiffness $k$ and the switch-on rate $r_0$ are in the vicinity of $k_c$ and $r_c$, respectively. {\bf Middle:} Below $k_c$, the MFPT always increases monotonously with the switch-off rate $r_1$, resulting in an optimal rate $r_1^*=0$. {\bf Right:} Above $k_c$, the MFPT is a non-monotonous function of $r_1$ and reaches a minimum at $r_1^*(k)>0$, provided $r_0$ is tuned near $r_c$.}
			\label{fig:Fig0}
\end{figure}

In a recent work\cite{Mercado_V_squez_2020}, we have exactly solved a model in one dimension that mimics stochastic resetting with an external potential. In this work, we studied a Brownian particle in a V-shaped piecewise linear potential with minimum at $X_0$, which was also taken as the initial position of the particle. The potential was turned on and off in a stochastic way at exponentially distributed times. 
The mean first passage time to a target placed at the origin in this model happens to be finite \cite{Mercado_V_squez_2020}. Furthermore, the MFPT can be optimized with a suitable choice of the switch-on and switch-off rates (that can be different), {\it i.e.}, by driving the particle out of the Boltzmann-Gibbs equilibrium. Intuitively, target search can be facilitated in the "off" state due to the absence of a potential barrier to overcome. Nevertheless, on the semi-infinite line, the potential has a beneficial effect as it prevents the particle from diffusing far away in the direction opposite to the target. The interplay between these two trends gives rise to a number of rather unexpected features which are not observed in simple resetting. In particular, taking the steady potential as a reference, a perturbative theory with the switch-off rate as a small parameter allowed us to expand the MFPT as:
\begin{equation}\label{eq0}
    MFPT=t^{(K)}(k)+r_1 f(k,r_0)+r_1^2g(k,r_0)+O(r_1^3),
\end{equation}
where $t^{(K)}$ is the Kramers' equilibrium time, $r_1$ the switch-off rate, $r_0$ the switch-on (resetting) rate, $k$ the potential stiffness, and $g(k,r_0)>0$. The coefficient $f(k,r_0)$ of the first order correction at small $r_1$, the so-called dispersion relation, is quite non-trivial and its sign tells us whether switching off/on the potential from time to time increases or decreases the equilibrium Kramers' time. Actually, a continuous phase transition in the optimal rate $r_1^*(k)$ which minimizes the MFPT occurs as the potential stiffness $k$ is varied. For $k<k_c$ (weak confinement regime), $f(k,r_0)>0$ for any $r_0$, therefore, from Eq. (\ref{eq0}), the optimal protocol consists in always keeping the potential on (or $r_1^*(k)=0$) and $r_0^*(k)$ is irrelevant. Conversely, for $k>k_c$ (strong confinement regime) there exists a window of values of $r_0$ centered around a certain $r_c$ such that $f(k,r_0)<0$, indicating that it is possible to improve the Kramers' time, as schematised in figure \ref{fig:Fig0}. The analysis of the marginal curve yields $k_c= 1.228780...$ in dimensionless units. At the transition, the dimensionless optimal switch-on rate is surprisingly large, $r_0^*(k_c)=r_c= 41.969 027...$, whereas for $k$ close to but above the critical value, the optimal switch-off rate grows continuously as $r_1^*(k)\simeq 32.913015\times(k-k_c)$ \cite{Mercado_V_squez_2020}. Meanwhile, the optimal MFPT decreases with $k$ and the classical resetting model of \cite{EvansPRL2011} is recovered in the limit $k\rightarrow\infty$ \cite{Mercado_V_squez_2020}.

In this paper, we extend these results by considering an arbitrary external potential applied intermittently and show that the above phenomenology drawn from the piecewise linear case qualitatively holds with other types of confining potentials. Experiments on resetting processes are typically performed in finite domains and with approximately harmonic optical traps \cite{besga2020optimal,tal2020experimental,besga2021dynamical,Faisant_2021}. We focus here on the harmonic case, but shall also consider more general confining potentials. Furthermore, our previous results, that were derived for a semi-infinite domain, are generalized to the case of finite intervals bounded by a reflective boundary.

This paper is organized as follows: in Section \ref{secMFPT} we introduce the model and the equations of motion that govern two quantities related to the distribution of the first passage time to a target site, namely, the survival probability in the Laplace domain and the MFPT. Although it seems difficult to solve these equations for a general external potential, in Section \ref{SecPerturI} we develop a perturbative method that allows us to exactly derive the dispersion relation $f(k,r_0)$ appearing in Eq. (\ref{eq0}) and which tells us under which conditions the intermittent potential improves the mean search time.  Section \ref{secPoteltialPower} is devoted to the analysis of a family of confining potentials of the form  $v(x)=k|x-1|^n/n$ in rescaled units. The harmonic case ($n=2$) and the linear potential ($n=1$) are treated separately in Sections \ref{SecHarmonicPot} and \ref{SecVPot}, respectively. The theoretical results obtained by the perturbative method are compared to those obtained from direct numerical integration of the coupled backward Fokker-Planck equations using a finite difference method. The latter method is also used to obtain the optimal rates beyond the perturbative regime ($k\gg k_c$). For completeness, in Section \ref{SecSteady} we derive the NESSs generated by intermittent harmonic potentials on the infinite line, in the absence of absorbing targets. These results are compared with Brownian dynamics simulations. Finally, in Section \ref{SecDiscussion} we conclude with a discussion.


\section{Mean First Passage Time}\label{secMFPT}

\begin{figure}[htp]
\centering
			\includegraphics[width=\textwidth]{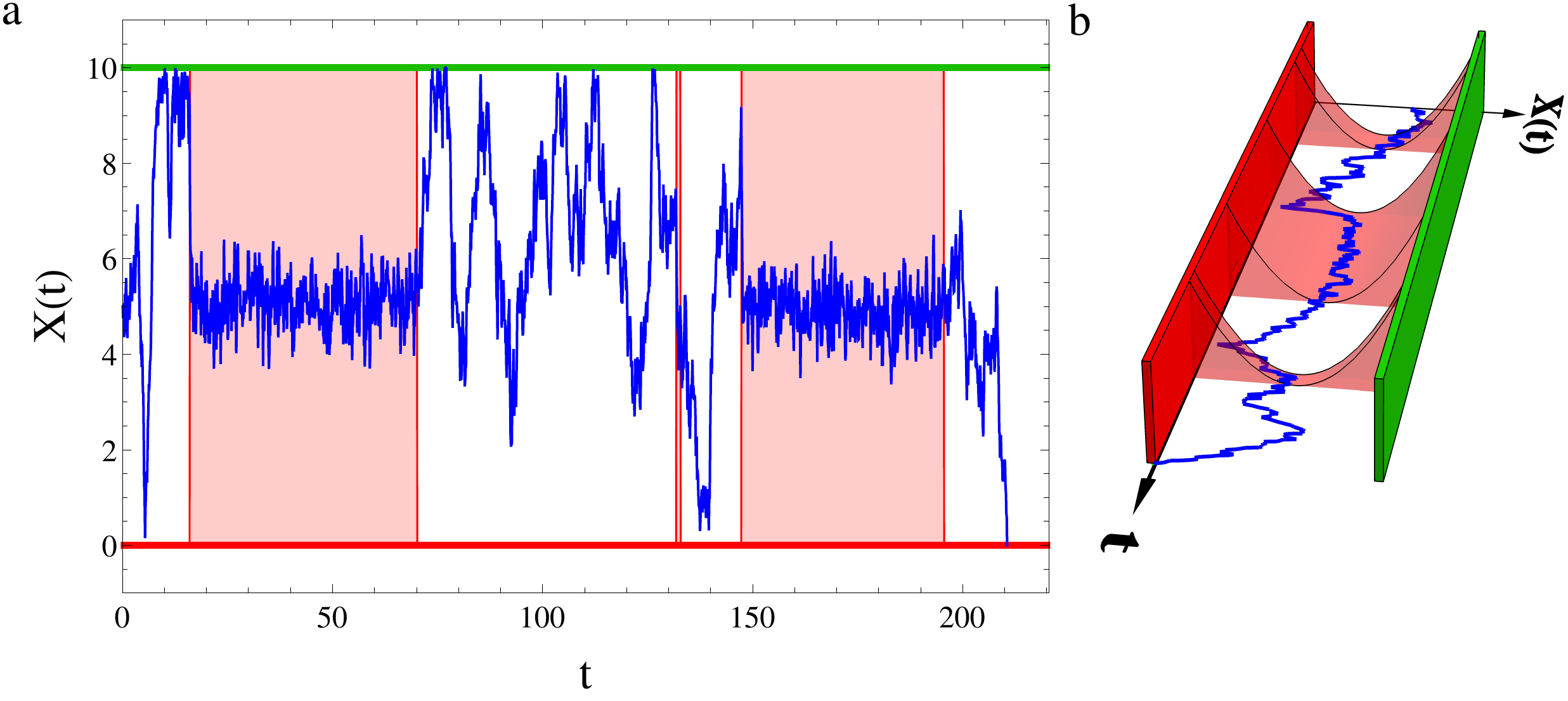}
            
			\caption{a) Trajectory of a diffusive particle with diffusion constant $D=1$, in an intermittent harmonic potential of the form $V(X)=\frac{K(X-5)^2}{2}$ with $K=5$. The shaded zones represent the time intervals when the potential is turned on (here $R_0=R_1=0.02$). An absorbing boundary is placed at $X=0$ (red line) and a reflective wall at $X=10$ (green line). When the particle is absorbed, the potential is either on or off. b) $3D$ view of a particle trajectory in the intermittent harmonic potential. }
			\label{fig:Fig1}
\end{figure}

In this Section we analyze the MFPT of a Brownian particle to an absorbing target site placed at the origin of a finite one-dimensional domain (see Fig. \ref{fig:Fig1}). In the domain, an external potential $V(X)$ is applied intermittently in time, so that the state of the potential is characterized by a binary function $\sigma(t)$ which takes the value $\sigma=0$ when the potential is switched off, and $\sigma=1$ when it is applied. The two-state process $\sigma(t)$ is characterized by two constant transition rates, $R_0$ (for the transition $0\rightarrow1$) and $R_1$ (for $1\rightarrow0$). The particle can always be absorbed by the target, independently of the state 0 or 1 of the potential. A reflecting wall is placed at the position $X=C$, with $C>0$. The semi-infinite domain case can be simply obtained by taking the limit $C\to \infty$. 

The evolution of the particle position $X(t)$ in the potential $\sigma(t)V(X)$ is given by the over-damped Langevin equation:
\begin{equation}
    \frac{dX(t)}{dt}= -\frac{1}{\Gamma}\sigma(t)V^{\prime}[X(t)]+\sqrt{2D}\xi(t),
    \label{ruleX1}
\end{equation}
where $\Gamma$ is the friction coefficient of the particle, $\xi(t)$ a Gaussian white noise of zero mean and correlations $\langle \xi(t)\xi(t')\rangle=\delta(t-t')$, and $D= k_BT/\Gamma$ the diffusion constant.

In the following, we will use the dimensionless space and time variables $x=X/X_0$ and $t/(X_0^2/D)$ (which we re-note as $t$), where $X_0$ is the distance between the minimum of $V(X)$ and the target placed at $X=0$. Let us define the dimensionless parameters:
\begin{eqnarray}
r_0&=&R_0X_0^2/D, \label{adimr0}\\ 
r_1&=&R_1X_0^2/D, \label{adimr1}
\end{eqnarray}
which are the rescaled \lq\lq on\rq\rq$\ $ and \lq\lq off\rq\rq $\ $ rates, respectively. The rescaled potential is given by $v(x)=V(xX_0)/(k_BT)$. The reflecting wall is placed at $x=c$ with $c=C/X_0>1$. In these dimensionless variables, the Langevin equation (\ref{ruleX1}) reduces to
\begin{equation}
\frac{dx}{dt}= - \sigma(t) v'(x) + \sqrt{2}\eta(t)
\end{equation}
where $\langle \eta(t)\rangle=0$ and $\langle \eta(t)\eta(t')\rangle= \delta(t-t')$.

Let us define $Q_0(x,t)$ as the probability that the particle has not hit the origin up to time $t$, given an initial position $x>0$ and initial potential state $\sigma(t=0)=0$. Similarly, $Q_1(x, t)$ corresponds to a
potential initially on. These survival probabilities satisfy the backward Fokker–Planck equations\cite{Mercado_V_squez_2020}
 \begin{eqnarray}
      \frac{\partial Q_1}{\partial t}=\frac{\partial^2 Q_1}{\partial x^2}- v^{\prime}(x)\frac{\partial Q_1}{\partial x} +r_1(Q_0-Q_1),\label{FPQP1}\\
      \frac{\partial Q_0}{\partial t}=\frac{\partial^2 Q_0}{\partial x^2}+r_0(Q_1-Q_0),\label{FPQP2}
\end{eqnarray}
valid in the domain $x\in[0,c]$. These expressions are obtained by extending the well-known steady potential case \cite{majumdar2007brownian,bray2013persistence} to a switching process with rates $r_0$ and $r_1$ \cite{benichou2011intermittent,mercado2019first}. 
Defining the Laplace transform $\widetilde{Q}(x,s)=\int^\infty_0 e^{-st}Q(x,t)dt$, Eqs. (\ref{FPQP1})-(\ref{FPQP2}) become
 \begin{eqnarray}
      -1=\frac{\partial^2 \widetilde{Q}_1}{\partial x^2}- v'(x)\frac{\partial \widetilde{Q}_1}{\partial x}-(s+r_1)\widetilde{Q}_1 +r_1\widetilde{Q}_0,\label{FPQP1s}\\
      -1=\frac{\partial^2 \widetilde{Q}_0}{\partial x^2}-(s+r_0)\widetilde{Q}_0+r_0\widetilde{Q}_1,\label{FPQP2s}
\end{eqnarray}
from which we deduce the relations for the corresponding MFPTs $t_0(x)$ and $t_1(x)$:
 \begin{eqnarray}
      -1=\frac{\partial^2 t_1(x)}{\partial x^2}- v'(x)\frac{\partial t_1(x)}{\partial x}-r_1\left[t_1(x) -t_0(x)\right],\label{TP1}\\
    -1=\frac{\partial^2 t_0(x)}{\partial x^2}-r_0\left[t_0(x)-t_1(x)\right],\label{TP2}
\end{eqnarray}
where we have used the usual relation $t_{\sigma}(x)=\widetilde{Q}_{\sigma}(x,s=0)$. 
The functions $t_0(x)$ and $t_1(x)$ will satisfy the following boundary conditions
\begin{eqnarray}
    t_{\sigma}(x=0)&=0,\label{BCts}\\
    \frac{\partial t_{\sigma}(x)}{\partial x}\Big|_{x=c}&=0,\label{BCdts}
\end{eqnarray}
where the initial state of the potential is $\sigma=\{0,1\}$. The first relation enforces the absorption at $x=0$, whereas the second one follows from imposing a zero flux through the reflective wall placed at $x=c$ \cite{gardiner2004handbook}.

Let us introduce the parameter
\begin{equation}
    \epsilon\equiv \frac{r_1}{r_0},
\end{equation}
which does not need to be small for the time being, and let us define the function
\begin{equation}\label{defS}
    S(x)\equiv t_0(x)-t_1(x).
\end{equation}
Eqs. (\ref{TP1})-(\ref{TP2}) can be rewritten as
 \begin{eqnarray}
     \frac{\partial^2 t_1(x)}{\partial x^2}-v'(x)\frac{\partial t_1(x)}{\partial x}=-1-\epsilon r_0S(x),\label{TP1S}\\
    \frac{\partial^2 S(x)}{\partial x^2}-r_0S(x)=-1-\frac{\partial^2 t_1(x)}{\partial x^2}.\label{TP2S}
\end{eqnarray}
By construction, $S(x)$  satisfies the boundary conditions
\begin{eqnarray}
    S(x=0)&=0,\label{BCS}\\
    \frac{\partial S(x)}{\partial x}\Big|_{x=c}&=0.\label{BCdS}
\end{eqnarray}
Eqs. (\ref{TP1S})-(\ref{TP2S}) take the following forms
 \begin{eqnarray}
     e^{v(x)}\frac{\partial}{\partial x}\left(e^{-v(x)}\frac{\partial}{\partial x}t_1(x)\right)=-1-\epsilon r_0S(x),\label{TP1S2}\\
     e^{\sqrt{r_0}x}\frac{\partial}{\partial x}\left(e^{-2\sqrt{r_0}x}\frac{\partial}{\partial x}e^{\sqrt{r_0}x}S(x)\right)=-1-\frac{\partial^2 t_1(x)}{\partial x^2}.\label{TP2S2}
\end{eqnarray}
Each equation can be integrated directly and one obtains coupled expressions for the general solutions $t_1(x)$ and $S(x)$:
\begin{eqnarray}
\fl
    t_1(x)=C_1\int^x_0 d\tau\ e^{v(\tau)}+C_2-\int^x_0 dy\ e^{v(y)}\int^y_0 dz\ e^{-v(z)}\left[1+\epsilon r_0S(z)\right]\label{t1integral},\\
\fl
    S(x)=C_3e^{-\sqrt{r_0}x}+C_4e^{\sqrt{r_0}x}+\frac{1}{r_0}-e^{-\sqrt{r_0}x}\int^x_0 dy\ e^{2\sqrt{r_0}y}\int^y_0 dz\ e^{-\sqrt{r_0}z}\frac{\partial^2 t_1(z)}{\partial z^2}\label{Sintegral}
\end{eqnarray}
where the constants $C_i$ are determined from the boundary conditions. Integrating by parts, the double integral in Eq. (\ref{Sintegral}) simplifies to a single integral:
\begin{equation}
S(x)=C_3e^{-\sqrt{r_0}x}+C_4e^{\sqrt{r_0}x}+\frac{1}{r_0}-\int^x_0 dy\ \frac{\partial t_1(y)}{\partial y}\cosh{\sqrt{r_0}(x-y)}\label{Sintegral2}.
\end{equation}
The integral of the r.h.s. can be integrated by parts again and written in terms of $t_1$ instead of its derivative. However, for the numerical evaluation of these expressions, it is more convenient to keep Eq. (\ref{Sintegral2}), as we will see later.
Imposing the boundary conditions (\ref{BCts})-(\ref{BCdts}) or (\ref{BCS})-(\ref{BCdS}) one gets
\begin{eqnarray}
\fl    t_1(x)= &\int^x_{0}dy\ e^{v(y)}\int^c_ydz\ e^{-v(z)}+\epsilon r_0\int^x_{0}dy\ e^{v(y)}\int^c_ydz\ e^{-v(z)} S(z),\label{t&S}\\
\fl    S(x)= &\frac{1}{r_0}-\frac{\cosh{\sqrt{r_0}(c-x)}}{r_0\cosh{\sqrt{r_0}c}}+\sinh{\sqrt{r_0}x}\int^c_0 dy \frac{\partial t_1(y)}{\partial y}\frac{\sinh{\sqrt{r_0}(c-y)}}{\cosh{\sqrt{r_0}c}}\nonumber\\
\fl &-\int^x_0 dy  \frac{\partial t_1(y)}{\partial y}\cosh{\sqrt{r_0}(x-y)}.\label{S&F&t}
\end{eqnarray}
Up to this point we have not made any approximation. Although we have obtained a formal solution of the system (\ref{TP1S})-(\ref{TP2S}), the functions are still coupled and {\it a priori} difficult to solve explicitly.

\section{Perturbative method for a general potential}\label{SecPerturI}

To make some progress, we develop an exact perturbative theory by expanding $S$ and $t_1$ in powers of $\epsilon=r_1/r_0$,
assuming $r_1\ll r_0$.
Let us look for solutions of the form:
\begin{eqnarray}
t_1(x)&=&t_1^{(0)}(x)+\epsilon t_1^{(1)}(x)+\epsilon^2 t_1^{(2)}(x)+\dots. \label{sert}\\ S(x)&=&S^{(0)}(x)+\epsilon S^{(1)}(x)+\epsilon^2 S^{(2)}(x)+\dots.\label{serS}
\end{eqnarray}
at small $\epsilon$. The function $t_1^{(0)}$ is related to the classic Kramers' problem of first passage over a steady potential barrier.
The functions $t_1^{(1)},\dots,S^{(0)},S^{(1)},\dots,$ depend on both the potential shape and the rate $r_0$, and can be determined recursively. We will particularly focus on $t_1(x)$, the MFPT of the Brownian particle starting from $x$ with a potential initially applied, and on its first order coefficient $t_1^{(1)}(x)$. The dispersion relation introduced in Eq. (\ref{eq0}) is identified with
\begin{equation}
    f(k,r_0)\rightarrow\frac{t_1^{(1)}(x)}{r_0}.
\end{equation}
When the coefficient $t_1^{(1)}(x)$ changes sign, a transition between two qualitatively different behaviours occurs. If the potential is such that
\begin{equation}
t_1^{(1)}(x)>0\ {\rm for\ any}\ r_0, 
\end{equation}
then switching the potential off and on back and forth ({\it i.e.}, setting $\epsilon$ small but $>0$) will always result in delaying target encounter on average compared to the case with the potential permanently applied, or $\epsilon=0$. Conversely, if
\begin{equation}
t_1^{(1)}(x)<0\ {\rm for\ some\ values\ of\ }  r_0, 
\end{equation}
then the intermittent dynamics of the potential can help to shorten the mean search time, as in the example sketched in figure \ref{fig:Fig0} .

\subsection{Leading order in \texorpdfstring{$\epsilon$}{Lg}}
By inserting the expansions (\ref{sert})-(\ref{serS}) into (\ref{t&S})-(\ref{S&F&t}) we obtain 
at leading order:
\begin{eqnarray}
\fl
t^{(0)}_1(x)=&\int^x_{0}dy \int^c_ydz\ e^{v(y)-v(z)},\label{TP1S0}\\
\fl
S^{(0)}(x)=&\frac{1}{r_0}-\frac{\cosh{\sqrt{r_0}(c-x)}}{r_0\cosh{\sqrt{r_0}c}}+\sinh{\sqrt{r_0}x}\int^c_0 dy \int^c_ydz\ e^{v(y)-v(z)}\frac{\sinh{\sqrt{r_0}(c-y)}}{\cosh{\sqrt{r_0}c}}\nonumber\\
\fl &-\int^x_0 dy  \int^c_y dz\ e^{v(y)-v(z)}\cosh{\sqrt{r_0}(x-y)}.\label{TP2S0}
\end{eqnarray}
The  solution of $t_1^{(0)}(x)$ in Eq. (\ref{TP1S0}) corresponds to the MFPT of the standard problem for a particle in a steady external potential $v(x)$ \cite{gardiner2004handbook}. It is related to the well-known Kramers' escape problem in equilibrium.  Eqs. (\ref{TP1S0})-(\ref{TP2S0}) combined with the definition (\ref{defS}) yield $t_0^{(0)}(x)$, which corresponds physically to the MFPT to the origin of the particle starting at $x$, with the potential initially "off" and which transits only once to the "on" state at rate $r_0$.

\subsection{Higher orders} 

At linear order in $\epsilon$, one obtains the aforementioned dispersion relation, one of the main results of this paper: 
\begin{eqnarray}
\fl    t_1^{(1)}(x)=&t_1^{(0)}(x)-\int^x_{0}dy \int^c_ydz\ e^{v(y)-v(z)}\frac{\cosh{\sqrt{r_0}(c-z)}}{\cosh{\sqrt{r_0}c}}\nonumber\\
\fl    &+r_0\left[\int^c_0 dy \int^c_y dz\ e^{v(y)-v(z)}\frac{\sinh{\sqrt{r_0}(c-y)}}{\cosh{\sqrt{r_0}c}}\right]\left[\int^x_{0}dy\int^c_ydz\ e^{v(y)-v(z)}\sinh{\sqrt{r_0}z}\right]\nonumber\\
\fl    &-r_0\int^x_{0}dy\int^c_ydz\int^z_0 du \int^c_u dw\ e^{v(y)-v(z)+v(u)-v(w)}\cosh{\sqrt{r_0}(z-u)},\label{TP1S1}
\end{eqnarray}
and
\begin{eqnarray}
    S^{(1)}(x)=&\sinh{\sqrt{r_0}x}\int^c_0 dy\ \frac{\partial t^{(1)}_1(y)}{\partial y}\frac{\sinh{\sqrt{r_0}(c-y)}}{\cosh{\sqrt{r_0}c}}\nonumber\\
 &-\int^x_0 dy\  \frac{\partial t_1^{(1)}(y)}{\partial y}\cosh{\sqrt{r_0}(x-y)}.\label{TP2S1} 
\end{eqnarray}

At order $\epsilon^m$, with $m$ an integer larger than one, Eqs. (\ref{t&S})-(\ref{S&F&t}) lead to
\begin{eqnarray}
    t_1^{(m)}(x)=&r_0\int^x_{0}dy\ e^{v(y)}\int^c_ydz\ e^{-v(z)}S^{(m-1)}(z),\label{TP1Sn}\\
     S^{(m)}(x)=&\sinh{\sqrt{r_0}x}\int^c_0 dy\ \frac{\partial t^{(m)}_1(y)}{\partial y}\frac{\sinh{\sqrt{r_0}(c-y)}}{\cosh{\sqrt{r_0}c}}\nonumber\\
  &-\int^x_0 dy\  \frac{\partial t_1^{(m)}(y)}{\partial y}\cosh{\sqrt{r_0}(x-y)},\label{TP2Sn}
\end{eqnarray}
The relatively simple relation between $t_1^{(n)}$ and $S_1^{(n-1)}$ allows us to recursively compute any $n$-th order term in principle. The expressions rapidly become complicated, though, and we will limit our analysis to the terms of order $\epsilon$, which are sufficient for our purpose.

\subsection{Semi-infinite line}

The above expressions can be written for the case of the semi-infinite line. This can be achieved by letting the position $c$ of the reflective wall tend to infinity. The expressions (\ref{TP1S0})-(\ref{TP2S0}) become
\begin{eqnarray}
t^{(0)}_1(x)=&\int^x_{0}dy\ \int^\infty_ydz\ e^{v(y)-v(z)},\label{TP1S0Infty}\\
S^{(0)}(x)=&\frac{1-e^{-\sqrt{r_0}x}}{r_0}+\sinh{\sqrt{r_0}x}\int^\infty_0 dy \int^\infty_y dz\ e^{v(y)-v(z)}e^{-\sqrt{r_0}y}\nonumber\\
&-\int^x_0 dy  \int^\infty_ydz\ e^{v(y)-v(z)}\cosh{\sqrt{r_0}(x-y)},\label{TP2S0Infty}
\end{eqnarray}
and at order one,
\begin{eqnarray}
\fl    t_1^{(1)}(x)=&t_1^{(0)}(x)-\int^x_{0}dy \int^\infty_ydz\ e^{v(y)-v(z)-\sqrt{r_0}z}\nonumber\\
\fl    &+r_0\left[\int^\infty_0 dy \int^\infty_y dz\ e^{v(y)-\sqrt{r_0}y-v(z)}\right]\left[\int^x_{0}dy\int^\infty_ydz\ e^{v(y)-v(z)}\sinh{\sqrt{r_0}z}\right]\nonumber\\
\fl    &-r_0\int^x_{0}dy\int^\infty_ydz\int^z_0 du \int^\infty_u dw\ e^{v(y)-v(z)+v(u)-v(w)}\cosh{\sqrt{r_0}(z-u)},\label{TP1S1Infinity}\\
\fl    S^{(1)}(x)=&\sinh{\sqrt{r_0}x}\int^\infty_0 dy\ \frac{\partial t^{(1)}_1(y)}{\partial y}e^{-\sqrt{r_0}y}-\int^x_0 dy\  \frac{\partial t_1^{(1)}(y)}{\partial y}\cosh{\sqrt{r_0}(x-y)},\label{TP2S1Infinity}
\end{eqnarray}

In the following we proceed to analyse these expressions, as well as Eqs. (\ref{TP1S0})-(\ref{TP2S1}) with a finite domain size $c$, for different potential shapes.

\section{Application to potentials of the form \texorpdfstring{$v(x)=k|x-1|^n/n$}{Lg}}\label{secPoteltialPower}
Let us consider a symmetric confining potential of the form $V(X)=K|X-X_0|^n/n$, with stiffness $K>0$ and exponent $n>0$, or $v(x)=k|x-1|^n/n$ in adimensional units. As the target is located at the origin and the minimum of the potential at $X_0$, in the dimensionless units defined in Section \ref{secMFPT} this minimum is at a distance unity from the target. The dimensionless stiffness is given by
\begin{equation}
    k=\frac{KX_0^n}{\Gamma D}.
\end{equation}
Although we obtained expressions for any value of the starting position $x$, we will focus on the case $x=1$, {\it i.e.}, the particle starting at the minimum of the potential. The particular case of the V-shaped potential ($n=1$)  has been analyzed in \cite{Mercado_V_squez_2020} on the semi-infinite line ($c=\infty$) by direct resolution of the first passage equations (\ref{TP1})-(\ref{TP2}).
In the following, let us define the function
\begin{equation}
\fl    G_n(x,c)=e^{\frac{k}{n}|x-1|^n}\int^c_x dy\ e^{-\frac{k}{n}|y-1|^n}=e^{\frac{k}{n}|x-1|^n}\frac{\gamma_{\frac{1}{n}}\left(\frac{k}{n}(c-1)^n\right)-\frac{|x-1|}{x-1}\gamma_{\frac{1}{n}}\left(\frac{k}{n}|x-1|^n\right)}{n^{1-\frac{1}{n}} k^{\frac{1}{n}}},\label{Gfun}
\end{equation}
where $\gamma_a(x)=\int^x_0 dz\ z^{a-1}e^{-z}$ is the lower incomplete gamma function. Inserting $v(x)=k|x-1|^n/n$ into Eq. (\ref{TP1S0}) gives
\begin{equation}
     t^{(0)}_1(x)=\int^x_0 dy\ G_n(y,c).\label{t10Vabs}
\end{equation}
Substituting the above expression into Eq. (\ref{TP2S0}) we obtain
\begin{eqnarray}
\fl    S^{(0)}(x)=&\frac{1}{r_0}-\frac{\cosh{\sqrt{r_0}(c-x)}}{r_0\cosh{\sqrt{r_0}c}}-\int^x_0 dy\  G_n(y,c)\cosh{\sqrt{r_0}(x-y)}\nonumber\\
\fl    &+\sinh{\sqrt{r_0}x}\int^c_0 dy\ G_n(y,c)\frac{\sinh{\sqrt{r_0}(c-y)}}{\cosh{\sqrt{r_0}c}}.\label{S0Vabs}
\end{eqnarray}
From Eq. (\ref{TP1S1}), the dispersion relation is re-expressed as
\begin{eqnarray}
\fl    t_1^{(1)}(x)=&t_1^{(0)}(x)-r_0\int^x_{0}dy\int^c_ydz\int^z_0 du \int^c_u dw\ e^{\frac{k}{n}\left(|y-1|^n-|z-1|^n+|u-1|^n-|w-1|^n\right)}\cosh{\sqrt{r_0}(z-u)}\nonumber\\
\fl    &-\int^x_{0}dy \int^c_ydz\ e^{\frac{k}{n}|y-1|^n-\frac{k}{n}|z-1|^n}\Bigg(\frac{\cosh{\sqrt{r_0}(c-z)}}{\cosh{\sqrt{r_0}c}}\nonumber\\
\fl    &-r_0\sinh{\sqrt{r_0}z}\left[\int^c_0 du\ G_n(u,c)\frac{\sinh{\sqrt{r_0}(c-u)}}{\cosh{\sqrt{r_0}c}}\right]\bigg).\label{t11npot}
\end{eqnarray}

\subsection{Harmonic potentials}\label{SecHarmonicPot}

Let us consider the important case $n=2$, {\it i.e.}, in which the switching potential is harmonic, or $v(x)=k(x-1)^2/2$.
When the potential is permanently applied, the particle follows a bounded Orstein-Uhlenbeck (OU) process of unit mean \cite{risken1984fokker}. The MFPT to the origin, starting from $x$, is given by the Kramers' relation  
\begin{equation}
    t^{(0)}_1(x)=\int^x_{0}dy\ e^{\frac{k}{2}(y-1)^2}\int^c_ydz\ e^{-\frac{k}{2}(z-1)^2}.\label{OUmfpt}
\end{equation}
(See \ref{OU section} for a standard derivation of this expression.)
As mentioned earlier, all the numerical results below will consider the starting position $x=1$.

Before discussing the effects of the off-on dynamics of the potential, one can notice from the above expression that $t_1^{(0)}(x=1)$ increases monotonically with $k$ when the position of the reflecting wall $c$ is below a particular value $c_0$. In this case, the MFPT is thus minimal at $k=0$, {\it i.e.}, when the particle diffuses freely without any external force and $t_1^{(0)}(k=0,c)=c-\frac{1}{2}$. On the other hand, when $c>c_0$, the MFPT $t_1^{(0)}$ exhibits a non-monotonic behaviour with $k$ and reaches a minimum at a certain value $k_{OU}>0$ which depends on $c$, see further the blue curve of Fig. \ref{fig:FigH2}a. The value of $c_0$ corresponds to the precise point in which the slope of $t_1^{(0)}(k,c)$ at $k=0$ changes from positive to negative values, \textit{i.e.},
\begin{equation}
    \frac{\partial t_1^{(0)}(k,c_0)}{\partial k}\Bigg |_{k=0}=0.
\end{equation}
Solving the above relation for $c_0$ using Eq. (\ref{OUmfpt}) we obtain 
\begin{equation}
    c_0=2.19148\dots
\end{equation}
This transition was studied in details in \cite{mercado2022freezing}.
Here, we will assume that the domain size is sufficiently large, or $c>c_0$, and will discuss the case $c<c_0$ afterwards.
Taking $n=2$ in Eq. (\ref{Gfun}), the function $G_2(x,c)$ can be recast as
\begin{equation}
\fl    G_2(x,c)=e^{\frac{k}{2}(x-1)^2}\sqrt{\frac{\pi}{2k}}\left[\mathrm{erf}\left( \sqrt{k/2}(c-1)\right)-\mathrm{erf}\left( \sqrt{k/2}(x-1)\right)\right],
\end{equation}
where $\mathrm{erf}(x)=\frac{2}{\sqrt{\pi}}\int^z_0dz\ e^{-z^2}$ is the error function. By replacing $G_2(x,c)$ into Eq. (\ref{t11npot}), $t_1^{(1)}(x)$ can be computed  by numerical integration.

\begin{figure}[t]
\centering
			\includegraphics[width=\textwidth]{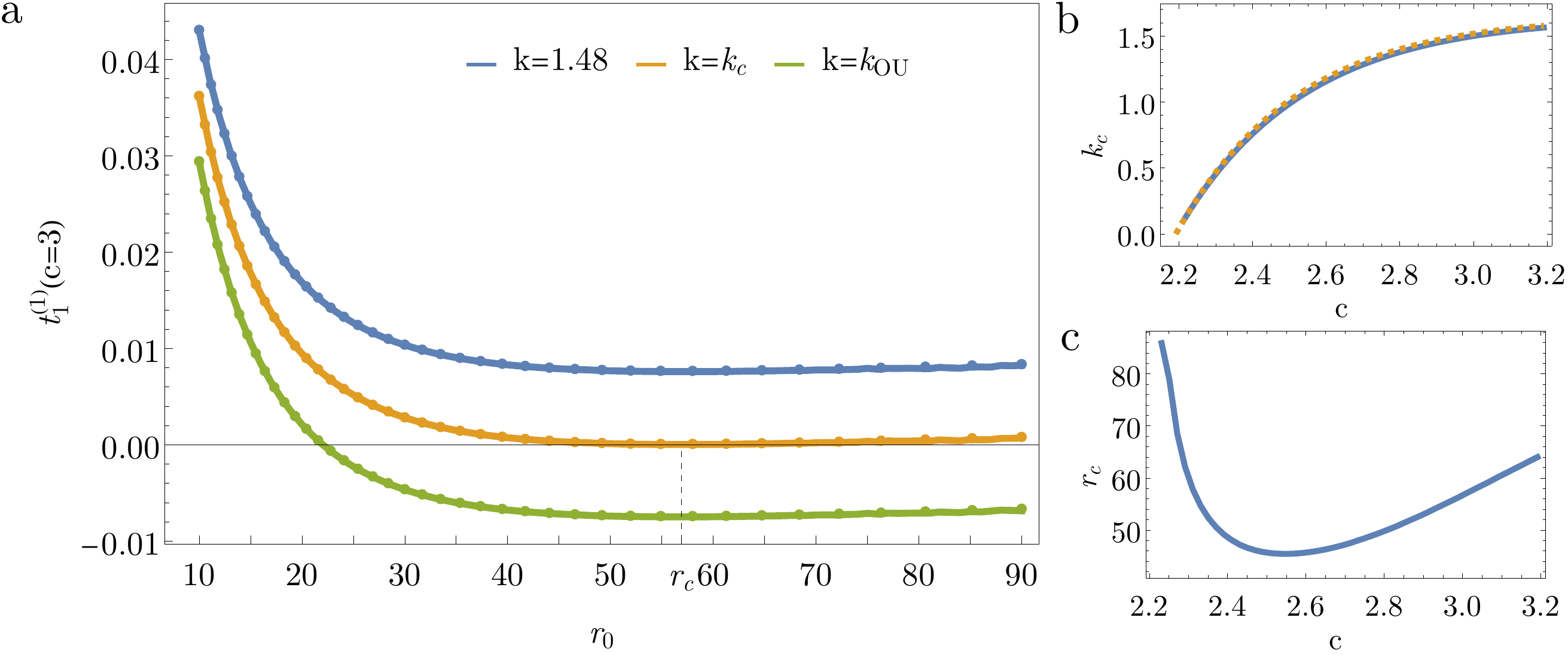}
			\caption{Searches starting from $x=1$ and with the harmonic potential on at $t=0$.  (a) Coefficient of the first correction in the series expansion of $t_1$ near $r_1=0$ (the \lq\lq dispertion relation\rq\rq) as a function of $r_0$ for a fixed domain size $c=3$, for various $k$ near $k_c(c=3)=1.49819\dots$. The points represent the results from numerically solving Eqs. (\ref{TP1})-(\ref{TP2}) by using a finite differences scheme.  (b) and (c): Critical potential stiffness $k_c$ (next to $k_{OU}$ shown with the dotted line) and optimal switch-off rate $r_c$ as a function of the domain size $c$, respectively. }
			\label{fig:FigH1}
\end{figure}

\begin{figure}[t]
\centering
			\includegraphics[width=\textwidth]{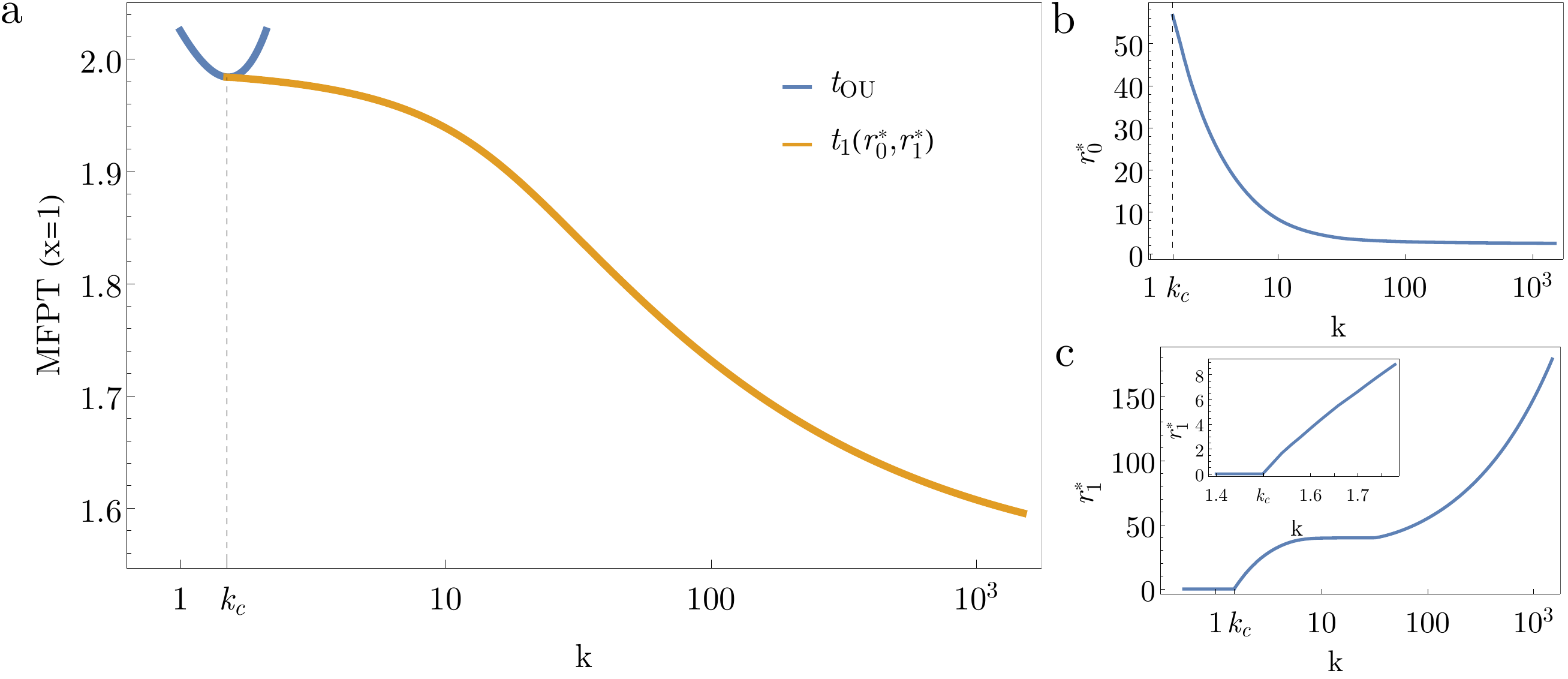}
			\caption{Searches starting from $x=1$ and with the harmonic potential on at $t=0$ and with $c=3$.  (a) Minimal MFPT $t_1^*$ as a function of the potential stiffness $k$. The orange line is obtained from numerical minimization of the numerical solution with respect to $(r_0,r_1)$. The blue line represents $t_1$ for a particle in a steady potential ($r_1=0$). (b) and (c): optimal rates $r_1^*(k)$ and $r_0^*(k)$. }
			\label{fig:FigH2}
\end{figure}

Figure \ref{fig:FigH1}a displays the dispersion relation as a function of $r_0$, obtained from evaluating $t_1^{(1)}(x)$ at $x=1$, in a domain of size $c=3$ and fixing $k$ at different representative values. Similarly to the scheme of figure \ref{fig:Fig0}, this function is non-monotonic with $r_0$. When the potential stiffness is below a certain critical value, or $k<k_c(c)=1.49823\dots$ for $c=3$, the function $t_1^{(1)}$ always stays positive. This means that turning alternatively the potential off (at a small rate $r_1$) and on (at any rate $r_0$) will always increase the mean search time compared with the Kramers' case $r_1=0$. At the marginal case $k=k_c$, however, the curve of $t_1^{(1)}$ becomes tangent to the $x-$axis, at a critical resetting rate $r_0=r_c$. One finds  surprisingly large value, $r_c=56.66926\dots$ for $c=3$. If the potential stiffness is slightly above $k_c$, there exists a small window of values of $r_0$ around $r_c$ for which $t_1^{(1)}$ is negative. Therefore, setting $r_0\simeq r_c$ and a small $r_1>0$ will shorten the mean search time, as illustrated in figure \ref{fig:Fig0}. The agreement between the theory and a direct numerical solution of Eqs. (\ref{TP1})-(\ref{TP2}) is excellent. To compute the numerical solutions, we used an implicit finite difference scheme which is described in \ref{finitdif}.

Figures \ref{fig:FigH1}b and \ref{fig:FigH1}c display the behaviour of the critical parameters $k_c$ and $r_c$, respectively, as a function of the domain size. The variations of the optimal stiffness $k_{OU}$ in equilibrium are also shown.
It is quite remarkable that $k_c$ is always very close to $k_{OU}$, but a bit lower. For instance, for $c=3$, we find $k_{OU}=1.51603\dots$, to be compared to the value $k_c=1.49823\dots$ mentioned above. Another surprising property is that $r_c$ is consistently much larger than unity: in dimensional units, the value of $r_0$ that minimizes $t_1$ is thus much larger than the inverse diffusion time to the target.

For $c<c_0$, one has $k_{OU}=0$ and a marginal dispersion relation cannot be found, therefore the pair $(k_c,r_c)$ cannot be defined. Numerical investigations indicate that $t_1^{(1)}$ is negative for all values of $k$ and $r_0$ in this case.

The results of Fig. \ref{fig:FigH1}a demonstrate the existence in large enough domains of a phase transition in the optimal parameters $(r_0^*,r_1^*)$, {\it i.e.}, the rates $(r_0,r_1)$ that minimize $t_1$. For $k<k_c$, we have $r_1^*(k)=0$; whereas $r_1^*(k)>0$ for $k>k_c$. Likewise, $r_0^*(k)$ is not defined below $k_c$, while $r_0^*(k)\simeq r_c$ for $k$ in the vicinity of $k_c$ and above. Determining the behavior of $r_1^*(k)$ slightly above $k_c$ would require an expansion at the following order $\epsilon^2$ in Eqs. (\ref{sert})-(\ref{serS}), see also Eq. (\ref{eq0}). Assuming that $t_1^{(1)}\propto k_c-k$ at $r_0=r_c$ and that the second order coefficient $t_1^{(2)}$ is $>0$ for $(k,r_0)\simeq (k_c,r_c)$, the minimization of the MFPT with respect to $r_1$ gives
\begin{equation}\label{orderparam}
    r_1^*(k)\propto k-k_c,
\end{equation}
close to the transition and when $k>k_c$. Therefore the transition is continuous with an exponent $\beta=1$ for the "order parameter" $r_1^*$.

We can alternatively solve Eqs. (\ref{TP1})-(\ref{TP2}) at $k$ fixed with the discrete element method described in  \ref{finitdif}, and minimize the solution $t_1(x=1)$ numerically with respect to $(r_0,r_1)$. Figure \ref{fig:FigH2}a displays the corresponding optimal search time $t_1^*(k)$ obtained. This quantity decreases monotonously with $k$. Figure \ref{fig:FigH2}b-c shows the corresponding optimal rates $r_0^*$ and $r_1^*$ as a function of $k$. The continuous character of the transition predicted by Eq. (\ref{orderparam}) can be appreciated in figure \ref{fig:FigH2}c and its inset.
Well above $k_c$, the optimal time of Fig. \ref{fig:FigH2}a becomes much smaller than the Kramers' time $t_1^{(0)}(k)$, which typically keeps increasing exponentially with $k$. In the limit $k\rightarrow\infty$, one recovers the optimal solution of the problem of diffusion with instantaneous stochastic resetting 
to the starting position, where $t_1^*\rightarrow 1.5451...$ and $r_0^*\rightarrow 2.5396...$ \cite{EvansPRL2011}.
 In very steep potentials, at large $k$, the diffusive particle is brought back very rapidly to the potential minimum at $x_0=1$. Therefore, the potential must be switched off rapidly to allow further diffusion and $r_1^*\rightarrow\infty$.

\subsection{V-shaped potential}\label{SecVPot}
\begin{figure}[t]
\centering
			\includegraphics[width=\textwidth]{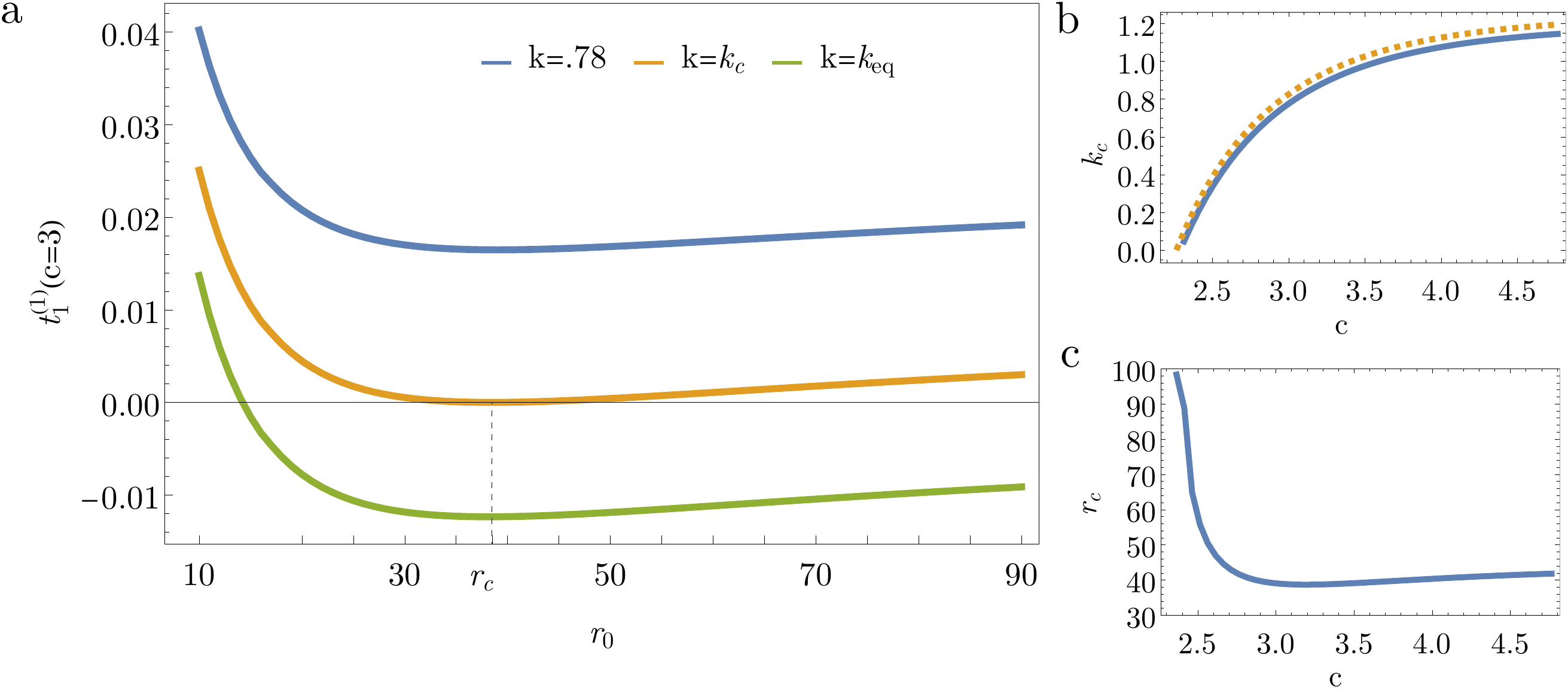}
			\caption{Searches starting from $x=1$ and with the potential on at $t=0$.  (a) Coefficient of the first correction in the series expansion of $t_1$ near $r_1=0$ (the \lq\lq dispertion relation\rq\rq) as a function of $r_0$ and at the fixed value $c=3$, for various $k$ near $k_c(c=3)=0.806777\dots$.  (b) and (c): Critical potential stiffness $k_c$ (next to $k_{OU}$ shown with the dotted line) and optimal switch-off rate $r_c$ as a function of the domains size $c$, respectively. }
			\label{fig:FigV1}
\end{figure}

In the case $n=1$, the switching potential is of the form $v(x)=k|x-1|$ and the rescaled potential strength given by
\begin{equation}
    k=\frac{KX_0}{\Gamma D}.
\end{equation}
The results are qualitatively similar to the harmonic case, see Fig. \ref{fig:FigV1} where the domain size is $c=3$, as well as to those obtained in \cite{Mercado_V_squez_2020} in the case of the semi-infinite line geometry or $c=\infty$. Notice that the curvature of the dispersion relation is larger near $r_c$ than in the harmonic case. 

Setting $n=1$ in the expression (\ref{Gfun}) for $G_n(x,c)$ we get
\begin{equation}
    G_1(x,c)=\frac{e^{k | x-1| }}{k} \left(1-e^{-k ( c-1) }-\frac{| x-1|}{x-1}\left(1-e^{-k | x-1|
   }\right)\right).
\end{equation}
Let us denote as $t_1^{(0)}(x,-)$ the solution in the range $0\le x<1$ and $t_1^{(0)}(x,+)$ the solution in $1<x<c$. Replacing the expression for $G_1(x,c)$ into Eq. (\ref{t10Vabs}) we obtain the Kramers' times
\begin{eqnarray}
    t_1^{(0)}(x,-)=&\frac{\left(2-e^{-k(c-1)}\right)\left(e^{k}-e^{-k (x-1)}\right)-xk}{k^2},\label{t10Vshape1}\\
    t_1^{(0)}(x,+)=&\frac{2(e^{k}-1)+(x-2)k+e^{-k(c-1)}(2-e^{k})-e^{-k (c-x)}}{k^2}\label{t10Vshape2}.
\end{eqnarray}

%
It is easy to check the continuity of the MFPT at $x=1$, or $t_1^{(0)}(1,-)=t_1^{(0)}(1,+)$.
The complete expressions of $S^{(0)}(x)$ and $t_1^{(1)}(x)$ are somehow intricate and we do not write them here. Taking $x=1$ in Eq. (\ref{t10Vshape1}) one gets 
\begin{equation}
t_1^{(0)}(k,c)=\frac{\left(2-e^{-k(c-1)}\right)\left(e^{k}-1\right)-k}{k^2},\label{t10Vshape1x1}
\end{equation}
where we have made explicit the dependency of the MFPT with the variables $k$ and $c$. 

As in the case of the harmonic potential, there exists a critical value $c_0$ such that $(i)$ if $c<c_0$ the minimum of the MFPT $t_1^{(0)}$ is achieved only at $k=0$ and, $(ii)$ if $c>c_0$, there is a finite potential strength $k$ at which the MFPT $t_1^{(0)}$ is minimum. The value $c_0$ satisfies 
\begin{equation}
    \frac{\partial t_1^{(0)}(k,c_0)}{\partial k}\Bigg |_{k=0}=0.
\end{equation}
Solving the above relation for $c_0$ and using Eq. (\ref{t10Vshape1x1}) we obtain 
\begin{equation}
    c_0=2.26376\dots
\end{equation}
If we let $c\to\infty$, the equations (\ref{t10Vshape1})-(\ref{t10Vshape2}) reduce to the simple form
\begin{eqnarray}
    t_1^{(0)}(x,-)=&\frac{2e^{k}\left(1-e^{-k x}\right)-xk}{k^2},\\
    t_1^{(0)}(x,+)=&\frac{2(e^{k}-1)+(x-2)k}{k^2},
\end{eqnarray}
which were obtained in \cite{Mercado_V_squez_2020}.
With these results and by using Eq. (\ref{TP2S0Infty}) we can calculate the leading order $S^{(0)}$ for the semi-infinite line,
\begin{eqnarray}
S^{(0)}(x,-)=&\frac{1-e^{-\sqrt{r_0}x}}{r_0}+\frac{2e^{k}\left(e^{-k x}-e^{-\sqrt{r_0}x}\right)}{k^2-r_0}+\frac{2e^{-\sqrt{r_0}}\sinh{\sqrt{r_0}x}}{\sqrt{r_0}(k+\sqrt{r_0})},\\
S^{(0)}(x,+)=&\frac{1-e^{-\sqrt{r_0}x}}{r_0}+\frac{2e^{-\sqrt{r_0}x}\left(\frac{k}{\sqrt{r_0}}\sinh{\sqrt{r_0}}+\cosh{\sqrt{r_0}}-e^{k}\right)}{k^2-r_0}.
\end{eqnarray}
Recalling that $t_0(x)=S(x)+t_1(x)$, deduce the first passage time with the initial condition $\sigma(t=0)=0$:
\begin{eqnarray}
\fl     t^{(0)}_0(x,-)=&\frac{1-e^{-\sqrt{r_0}x}}{r_0}+\frac{2e^{k}\left(1-e^{-k x}\right)-xk}{k^2}+\frac{2e^{k}\left(e^{-k x}-e^{-\sqrt{r_0}x}\right)}{k^2-r_0}\nonumber\\
\fl &+\frac{2e^{-\sqrt{r_0}}\sinh{\sqrt{r_0}x}}{\sqrt{r_0}(k+\sqrt{r_0})},\\
\fl    t^{(0)}_0(x,+)=&\frac{1-e^{-\sqrt{r_0}x}}{r_0}+\frac{2(e^{k}-1)+(x-2)k}{k^2}\nonumber\\
\fl &+\frac{2e^{-\sqrt{r_0}x}\left(\frac{k}{\sqrt{r_0}}\sinh{\sqrt{r_0}}+\cosh{\sqrt{r_0}}-e^{k}\right)}{k^2-r_0},
\end{eqnarray}
recovering the results obtained in \cite{Mercado_V_squez_2020}.

\section{Stationary density with an intermittent harmonic potential}\label{SecSteady}

In this Section we derive the exact expressions for the non-equilibrium stationary states generated by an intermittent harmonic potential on the infinite line. These results generalise the case $r_0=r_1=\gamma$ studied in \cite{santra2021brownian}. 
Let us introduce $P_\sigma(X,t)$ as the joint probability density that the particle is in the vicinity of $X$ and the potential in state $\sigma=\{0,1\}$ at time $t$ (the initial condition being implicit). The complete probability density is given by
\begin{equation}
    P(X,t)=P_0(X,t)+P_1(X,t),
\end{equation}
and we will focus on $\lim_{t\rightarrow\infty} P(X,t)$.

For a general intermittent potential $V(X)$, these densities satisfy the forward Fokker-Planck equations
\begin{eqnarray}
\fl     \frac{\partial}{\partial t}P_0(X,t)&=D\frac{\partial^2}{\partial X^2}P_0(X,t)-R_0 P_0(X,t)+R_1P_1(X,t),\label{dpst0}\\
\fl     \frac{\partial }{\partial t}P_1(X,t)&=D\frac{\partial^2}{\partial X^2}P_1(X,t)+\frac{1}{\Gamma}\frac{\partial}{\partial X}\left[ V^{\prime}(X)P_1(X,t)\right]-R_1 P_1(X,t)+R_0P_0(X,t)\label{dpst1}.
\end{eqnarray}
For a harmonic potential $V(X)=\frac{K}{2}X^2$ (we now place the minimum of the potential at the origin), Eqs. (\ref{dpst0})-(\ref{dpst1}) read
\begin{eqnarray}
\fl \frac{\partial}{\partial t}P_0(X,t)&=D\frac{\partial^2}{\partial X^2}P_0(X,t)-R_0 P_0(X,t)+R_1P_1(X,t),\label{dpst0p}\\
\fl     \frac{\partial }{\partial t}P_1(X,t)&=D\frac{\partial^2}{\partial X^2}P_1(X,t)+\frac{K}{\Gamma}\frac{\partial}{\partial X}\left[ XP_1(X,t)\right]-R_1 P_1(X,t)+R_0P_0(X,t)\label{dpst1p}.
\end{eqnarray}
We again employ the dimensionless variables and parameters $x=X/L$, $t/(L^2/D)$ (re-noted as $t$), $r_0=R_0L^2/D$, $r_1=R_1L^2/D$, where $L$ is an arbitrary length. The rescaled potential stiffness $k$ reads
\begin{eqnarray}
k&=&\frac{K L^2}{\Gamma D}, \label{adimK}
\end{eqnarray}
The joint densities associated to $x=X/L$ are denoted as $p_0(x,t)$ and $p_1(x,t)$.
In the steady state limit, the time derivatives are set to zero in Eqs. (\ref{dpst0p})-(\ref{dpst1p}), 
\begin{eqnarray}
\frac{\partial^2 p_0(x)}{\partial x^2}-r_0 p_0(x)+r_1p_1(x)=0,\label{dpst0p2}\\
\frac{\partial^2 p_1(x)}{\partial x^2 }+k\frac{\partial}{\partial x}\left[xp_1(x)\right]-r_1 p_1(x)+r_0p_0(x)=0\label{dpst1p2}.
\end{eqnarray}
Taking the spatial Fourier transform $\widetilde{f}(\nu)=\int_{-\infty}^\infty dx\ e^{-i\nu x}f(x)$ of  Eqs. (\ref{dpst0p2})-(\ref{dpst1p2}) gives,
\begin{eqnarray}
-\left(\nu^2+r_0 \right)\widetilde{p}_0(\nu)+r_1\widetilde{p}_1(\nu)=0,\label{dpst0p3}\\
-\left(\nu^2+r_1\right)\widetilde{p}_1(\nu)-k\nu \frac{\partial}{\partial\nu}\widetilde{p}_1(\nu)+r_0\widetilde{p}_0(\nu)=0\label{dpst1p3},
\end{eqnarray}
where we have use the identities $\widetilde{\frac{\partial f(x)}{\partial x}}=i\nu \widetilde{f}(\nu)$ and $\widetilde{xf(x)}=i \frac{\partial\widetilde{f}(\nu)}{\partial \nu}$.
Combining Eqs. (\ref{dpst0p3})-(\ref{dpst1p3}) gives
\begin{equation}
    \frac{\partial \widetilde{p}_1(\nu)}{\partial\nu}+\frac{\nu\left(\nu^2+r_0+r_1\right)}{k\left(\nu^2+r_0\right)}\widetilde{p}_1(\nu)=0,
\end{equation}
which is solved as
\begin{equation}
    \widetilde{p}_1(\nu)=Ae^{-\int^{\nu} a(\tau)d\tau},
\end{equation}
where
\begin{equation}
    a(\tau)=\frac{\tau\left(\tau^2+r_0+r_1\right)}{k\left(\tau^2+r_0\right)},\label{afunction}
\end{equation}
and $A$ is a constant to be determined later from the normalization condition. We obtain
\begin{equation}
    \widetilde{p}_1(\nu)=Ae^{-\frac{\nu^2+r_1\ln{\left(\nu^2+r_0\right)}}{2k}}=\frac{Ae^{-\frac{\nu^2}{2k}}}{\left(\nu^2+r_0\right)^{\frac{r_1}{2k}}}.\label{p1sigma}
\end{equation}
From Eq. (\ref{dpst0p3}) the density $\widetilde{p}_0(\nu)$ is given by
\begin{equation}
    \widetilde{p}_0(\nu)=\frac{r_1}{\nu^2+r_0}\widetilde{p}_1(\nu)=\frac{Ar_1e^{-\frac{\nu^2}{2k}}}{\left(\nu^2+r_0\right)^{\frac{r_1}{2k}+1}}.\label{p0sigma}
\end{equation}
The normalization condition imposes 
\begin{equation}
\widetilde{p}_0(\nu=0)+\widetilde{p}_1(\nu=0)=\int^{\infty}_{-\infty}dx\left[p_0(x)+p_1(x)\right]=1,  
\end{equation}
from which we deduce
\begin{equation}
    A=\frac{r_0^{\frac{r_1}{2k}+1}}{r_0+r_1}.
\end{equation}
The full position density $\widetilde{p}(\nu)=\widetilde{p}_0(\nu)+\widetilde{p}_1(\nu)$ therefore reads
\begin{equation}
    \widetilde{p}(\nu)=\frac{r_0^{\frac{r_1}{2k}+1}\left(\nu^2+r_0+r_1\right)e^{-\frac{\nu^2}{2k}}}{\left(r_0+r_1\right)\left(\nu^2+r_0\right)^{\frac{r_1}{2k}+1}}.\label{pofk}
\end{equation}
By setting $r_0=r_1\equiv\gamma$, this expression reduces to the result derived in \cite{santra2021brownian}.
We now consider a few limiting cases, where this expression simplifies.

\subsection{Limits \texorpdfstring{$r_0=\infty$}{Lg} or \texorpdfstring{$r_1=0$}{Lg}}
In those two cases, the potential always stays in the \lq\lq on" state and  Eq. (\ref{pofk}) reduces to
\begin{equation}
    \widetilde{p}(\nu,r_0=\infty,r_1=0)=e^{-\frac{\nu^2}{2k}},\label{pofkOH}
\end{equation}
which is easily inverted as
\begin{equation}
 p(x,r_0=\infty,r_1=0)=p_{OU}(x)=\sqrt{\frac{k}{2\pi}}e^{-\frac{kx^2}{2}}.\label{OrnsteinU}
\end{equation}
One recovers the equilibrium distribution $p_{OU}(x)$ for the Ornstein-Uhlenbeck process \cite{risken1984fokker}.

\subsection{Limit \texorpdfstring{$r_1\ll k$}{Lg}}
In this limit, the potential is steep or $r_1/k\approx 0$. The probabilities $\widetilde{p}_0$ and $\widetilde{p}_1$ in Eqs. (\ref{p1sigma})-(\ref{p0sigma}) take the approximate forms
\begin{eqnarray}
    \widetilde{p}_1(\nu)&=\frac{r_0}{r_0+r_1}e^{-\frac{\nu^2}{2k}},\label{p1sigma2}\\
    \widetilde{p}_0(\nu)&=\frac{r_1}{r_0+r_1}\left(\frac{r_0e^{-\frac{\nu^2}{2k}}}{\nu^2+r_0}\right).\label{p0sigma2}
\end{eqnarray}
The inverse Fourier transform of Eq. (\ref{p1sigma2}) reduces to the Ornstein-Uhlenbeck distribution (\ref{OrnsteinU}), weighted by the probability that the potential is turned on. On the other hand, the inverse transform of Eq. (\ref{p0sigma2})  can be obtained from the convolution theorem by noticing that the inverse transform of $r_0/(\nu^2+r_0)$ is $\frac{\sqrt{r_0}}{2}e^{-\sqrt{r_0}|x|}$, and the inverse transform of $e^{-\nu^2/2k}$ is again given by Eq. (\ref{OrnsteinU}). Therefore
\begin{eqnarray}
p_0(x)&=\frac{r_1}{r_0+r_1}\int^\infty_{-\infty}\left(\sqrt{\frac{k}{2\pi}}e^{-\frac{ky^2}{2}}\right)\frac{\sqrt{r_0}}{2}e^{-\sqrt{r_0}|x-y|}dy.\label{p0smallr1}
\end{eqnarray}
One recognizes in this result the probability distribution of an instantaneous resetting process with rate $r_0$, averaged over a equilibrium Orstein-Uhlenbeck distribution of resetting points, $p_{OU}(y)=\sqrt{\frac{k}{2\pi}}e^{-\frac{ky^2}{2}}$, which is itself weighted by the probability that the potential is turned off \cite{Evans_2011}. In the steep potential limit, the total probability density $p(x)=p_0(x)+p_1(x)$ thus reads
\begin{equation}\label{px1}
\fl    p(x)=\frac{r_0}{r_0+r_1}\sqrt{\frac{k}{2\pi}}e^{-\frac{kx^2}{2}}+\frac{r_1}{r_0+r_1}\int^\infty_{-\infty}\left(\sqrt{\frac{k}{2\pi}}e^{-\frac{ky^2}{2}}\right)\frac{\sqrt{r_0}}{2}e^{-\sqrt{r_0}|x-y|}dy.
\end{equation}
We rewrite the integral in Eq. (\ref{p0smallr1}) as
\begin{eqnarray}
\fl p_0(x)&=\frac{r_1}{2(r_0+r_1)}\sqrt{\frac{r_0k}{2\pi}}\left(\int^x_{-\infty}e^{-\frac{ky^2}{2}-\sqrt{r_0}(x-y)}dy+\int^\infty_{x}e^{-\frac{ky^2}{2}-\sqrt{r_0}(y-x)}dy\right)\nonumber\\
\fl &=\frac{r_1e^{\frac{r_0}{2k}}}{2(r_0+r_1)}\sqrt{\frac{r_0k}{2\pi}}\left(e^{-\sqrt{r_0}x}\int^x_{-\infty}e^{-\frac{k(y-\sqrt{r_0}/k)^2}{2}}dy+e^{\sqrt{r_0}x}\int^\infty_{x}e^{-\frac{k(y+\sqrt{r_0}/k)^2}{2}}dy\right)\nonumber\\
\fl &=\frac{r_1\sqrt{r_0}e^{\frac{r_0}{2k}}}{4(r_0+r_1)}\left(e^{-\sqrt{r_0}x}\mathrm{erfc}\left(\frac{\sqrt{r_0}-kx}{\sqrt{2k}}\right)+e^{\sqrt{r_0}x}\mathrm{erfc}\left(\frac{\sqrt{r_0}+kx}{\sqrt{2k}}\right)\right).\label{p0smallr1_2}
\end{eqnarray}{}
After Eq. (\ref{px1}) or (\ref{p0smallr1_2}), $p(x)$ is symmetric and in the asymptotic analysis below,
we consider the positive part, when $x\to +\infty$. The negative tail follows from symmetry. At large $z$, we use $\mathrm{erfc(z)}\approx e^{-z^2}/\sqrt{\pi}z$ and $\mathrm{erfc(-z)}\approx 2-e^{-z^2}/\sqrt{\pi}z$. One deduces the large $x$ behavior
\begin{equation}
\fl   p_0(x)\approx \frac{r_1\sqrt{r_0}e^{\frac{r_0}{2k}}}{2(r_0+r_1)}\left(e^{-\sqrt{r_0}x}-\frac{\sqrt{2r_0k}e^{-\frac{kx^2}{2}-\frac{r_0}{2k}}}{\sqrt{\pi}(k^2x^2-r_0)}\right)\approx \frac{r_1e^{\frac{r_0}{2k}}}{r_0+r_1}\left(\frac{\sqrt{r_0}e^{-\sqrt{r_0}x}}{2}\right).
\end{equation}
Thus $p_0(x)$ decays exponentially for large $x$, which is much slower than the Gaussian decay of $p_1(x)$. Adding the two and using symmetry, one obtains the total probability density $p(x)$ at large $|x|$
\begin{equation}
    p(x)\approx\frac{r_1\sqrt{r_0}e^{\frac{r_0}{2k}-\sqrt{r_0}|x|}}{2(r_0+r_1)}.\label{smallr1largex}
\end{equation}

\subsection{General case}
To tackle the general case for arbitrary rates $r_0$ and $r_1$, we take advantage of the convolution theorem again, noticing that the inverse Fourier transform of $(\nu^2+r_0)^{-a}$ is \cite{abramowitz}
\begin{equation}
\fl   \frac{1}{2\pi} \int^\infty_{-\infty}\frac{e^{i\nu x}d\nu}{\left(\nu^2+r_0\right)^a}=
     \frac{1}{\pi}\int^\infty_{0}\frac{\cos{(\nu x})d\nu}{\left(\nu^2+r_0\right)^a}=\frac{\left(2\sqrt{r_0}|x|^{-1}\right)^{\frac{1}{2}-a}}{\sqrt{\pi}\Gamma(a)}\mathbb{K}_{a-\frac{1}{2}}\left(\sqrt{r_0}|x|\right),
\end{equation}
where $\mathbb{K}_\alpha(x)$ is the modified Bessel function of the second kind and $\Gamma(\cdot)$ the Gamma function. Eq. (\ref{pofk}) can be recast as
\begin{eqnarray}
\fl   p(x)=&\frac{r_1}{r_0+r_1}\left[\frac{2^{-\frac{r_1+k}{2k}}r_0^{\frac{r_1+3k}{4k}}}{\sqrt{\pi}\Gamma(\frac{r_1}{2k}+1)}\right]\int^{\infty}_{-\infty} \left(\sqrt{\frac{k}{2\pi}}e^{-\frac{ky^2}{2}}\right)|x-y|^{\frac{r_1+k}{2k}}\mathbb{K}_{\frac{r_1+k}{2k}}\left(\sqrt{r_0}|x-y|\right)dy\nonumber\\
\fl    &+\frac{r_0}{r_0+r_1}\left[\frac{2^{-\frac{r_1-k}{2k}}r_0^{\frac{r_1+k}{4k}}}{\sqrt{\pi}\Gamma(\frac{r_1}{2k})}\right]\int^{\infty}_{-\infty} \left(\sqrt{\frac{k}{2\pi}}e^{-\frac{ky^2}{2}}\right)|x-y|^{\frac{r_1-k}{2k}}\mathbb{K}_{\frac{r_1-k}{2k}}\left(\sqrt{r_0}|x-y|\right)dy.
\end{eqnarray}
\begin{figure}[htp]
\centering
			\includegraphics[width=.6\textwidth]{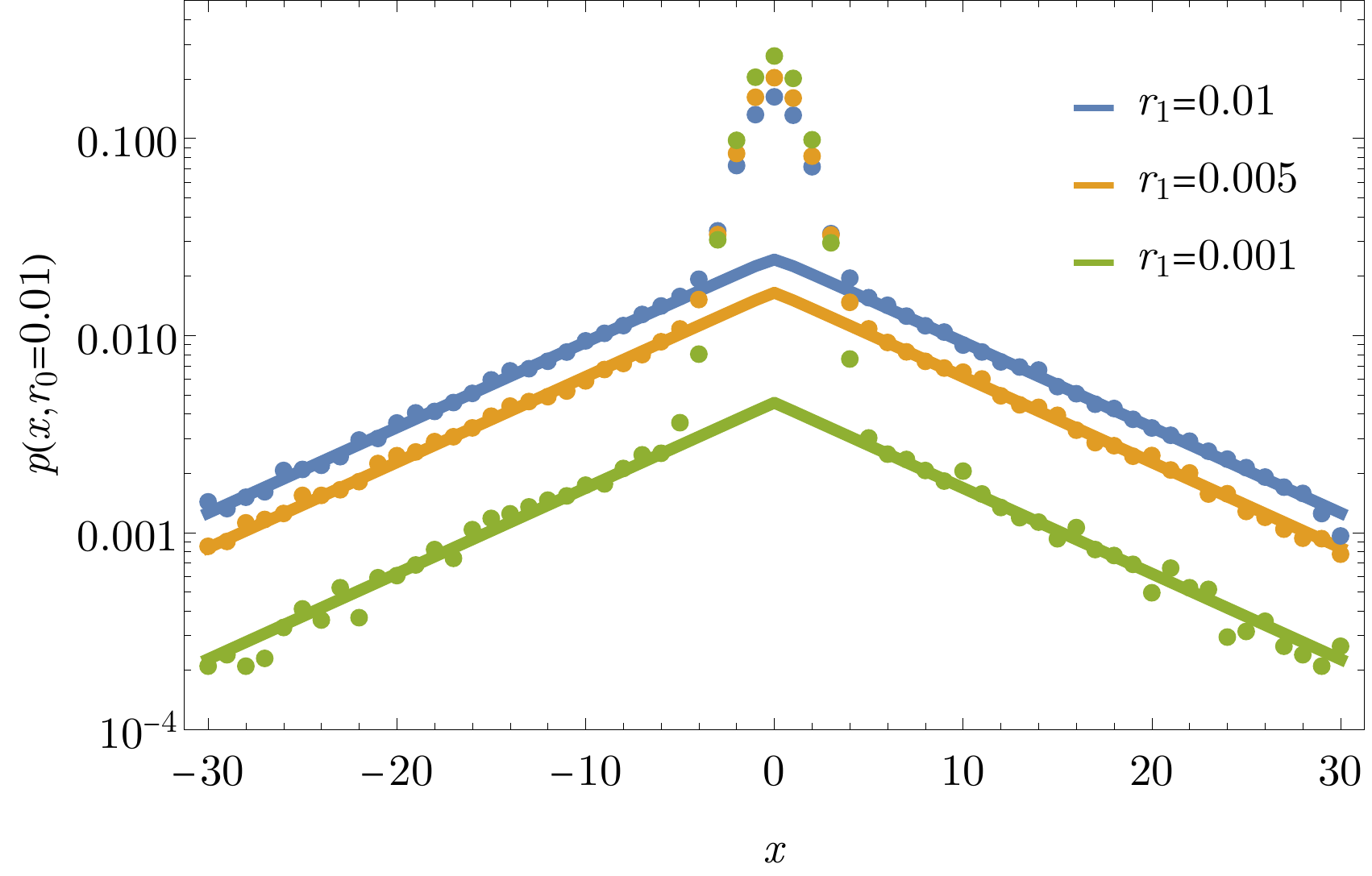}
            
			\caption{Particle density for an intermittent potential of the form $v(x)=\frac{k}{2}x^2$, with fixed rate $r_0=0.01$ and stiffness $k=1$, and for several values of $r_1$. The solid lines represent the approximation (\ref{plargex}) and the symbols simulation results obtained with the Gillespie algorithm.}
			\label{fig:RelativeVarianceL}
\end{figure}
The large $x$ behavior is obtained from the expansion $\mathbb{K}_{\alpha}(z)\approx \sqrt{\pi}e^{-z}/\sqrt{2z}$ at large $z$, or
\begin{eqnarray}
 \fl   &\int^{\infty}_{-\infty} \left(\sqrt{\frac{k}{2\pi}}e^{-\frac{ky^2}{2}}\right)|x-y|^{\frac{r_1+k}{2k}}\mathbb{K}_{\frac{r_1+k}{2k}}\left(\sqrt{r_0}|x-y|\right)dy\nonumber\\
\fl    &\approx \int^{\infty}_{-\infty} \left(\frac{\sqrt{k}e^{-\frac{ky^2}{2}}}{2}\right)r_0^{-\frac{1}{4}}(x-y)^{\frac{r_1}{2k}}e^{-\sqrt{r_0}(x-y)}dy\nonumber\\
\fl    &\approx r_0^{-\frac{1}{4}}x^{\frac{r_1}{2k}}e^{\frac{r_0}{2k}-\sqrt{r_0}x}\int^{\infty}_{-\infty} \left(\frac{\sqrt{k}e^{-\frac{k(y-\sqrt{r_0}/k)^2}{2}}}{2}\right)dy=\sqrt{\frac{\pi}{2}}r_0^{-\frac{1}{4}}x^{\frac{r_1}{2k}}e^{\frac{r_0}{2k}-\sqrt{r_0}x},\label{KBesselApp}
\end{eqnarray}
similarly, 
\begin{eqnarray}
\fl    &\int^{\infty}_{-\infty} \left(\sqrt{\frac{k}{2\pi}}e^{-\frac{y^2}{2}}\right)|x-y|^{\frac{r_1-k}{2k}}\mathbb{K}_{\frac{r_1-k}{2k}}\left(\sqrt{r_0}|x-y|\right)dy\approx
    \sqrt{\frac{\pi}{2}} r_0^{-\frac{1}{4}}x^{\frac{r_1}{2k}-1}e^{\frac{r_0}{2k}-\sqrt{r_0}x}.\label{KBesselApp2}
\end{eqnarray}
Combining the expressions (\ref{KBesselApp})-(\ref{KBesselApp2}) one obtains
\begin{eqnarray}
\fl   p(x)\approx\frac{r_0^{\frac{r_1+2k}{4k}}x^{\frac{r_1}{2k}}e^{\frac{r_0}{2k}-\sqrt{r_0}x}}{2^{\frac{r_1}{2k}}\Gamma(\frac{r_1}{2k})(r_0+r_1)}\left(k+\frac{\sqrt{r_0}}{x}\right)\approx\frac{k r_0^{\frac{r_1+2k}{4k}} e^{\frac{r_0}{2k}}}{2^{\frac{r_1}{2k}}\Gamma(\frac{r_1}{2k})(r_0+r_1)} |x|^{\frac{r_1}{2k}}e^{-\sqrt{r_0}|x|},\label{plargex}
\end{eqnarray}
where we have replaced $x$ by $|x|$ by symmetry.
We conclude that the exponential
tail of the non-equilibrium steady state distribution exhibits an algebraic prefactor, or $p(x)\sim |x|^{\frac{r_1}{2k}}e^{-\sqrt{r_0}|x|}$. The power-law exponent is non-trivial and depends continuously on the system parameters.
Notice that in the limit $r_1\approx 0$, $\Gamma(\frac{r_1}{2k})\approx \frac{2k
}{r_1}$ and Eq. (\ref{plargex}) reduces to Eq. (\ref{smallr1largex}). Our results are in very good agreement with simulation results obtained with the Gillespie algorithm\cite{GILLESPIE1976403}, as displayed in Fig. \ref{fig:RelativeVarianceL}.
 
\section{Discussion}\label{SecDiscussion}

In comparison with the classical Kramers' problem \cite{kramers1940brownian}, the theory of first passage times in time-dependent or fluctuating potentials is much less understood.
We have studied the motion of a Brownian particle which is driven out-of-equilibrium by means of an intermittent confining potential in a $1d$ bounded domain. The dynamics of the potential is defined by a two-state process, characterized by two rates $r_0$ and $r_1$. When the potential is "on", the particle is attracted toward the potential minimum, similarly to a resetting process. In the "off" state, the particle diffuses freely and has no barrier to overcome to reach an absorbing target. By properly choosing the rates $r_0$ and $r_1$, one can minimize the mean search time for a target located at a fixed position. The optimal choice of these parameters becomes non-trivial when the rescaled potential stiffness is larger than a critical value. In this case, the optimal switch-off rate $r_1^*$ is non-zero and the mean first passage time is improved with respect to the Kramers' time. Importantly, this transition exists if the domain size is large enough, the positions of the target and potential minimum being fixed. We have studied the dependence of the critical parameters on the domain size. In the limit of infinite stiffness, the particle undergoes instantaneous and perfect resetting to the potential minimum. In this case $r_1^*\rightarrow\infty$ and the well-studied resetting model introduced in \cite{EvansPRL2011} is recovered.

In unbounded domains free of absorbing targets and with an intermittent harmonic potential, we have shown that the particle always reaches a non-equilibrium stationary state, for any non-vanishing $r_0$ and $r_1$. Due to the intermittency of the potential, the particle density strongly differs from the Boltzmann-Gibbs distribution, but it does not take the simple exponential form of a pure resetting process either: it exhibits exponential tails with a power-law prefactor instead. One can actually notice from Eq. (\ref{plargex}) that this power-law correction disappears in the large stiffness limit. These findings are similar to those of \cite{santra2021brownian}, corresponding to the case $r_0=r_1=\gamma$, although the power-law corrections were not determined explicitly.

Our set-up is reminiscent of the phenomenon of resonant activation \cite{ResonantDoering1992,redner2001guide}. In that problem, the time taken by a Brownian particle to cross a fluctuating energy barrier (of a piecewise linear potential) is calculated. The barrier height switches stochastically between a higher and a lower value at some rate $\gamma$, which is the same for both transitions. Interestingly, the mean first crossing time can be minimized for a finite value of $\gamma$, which thus corresponds to a "resonance" for the reaction rate. Our set-up is somehow similar to the resonant activation set-up with a vanishing lower barrier (no potential), and if one sets $r_0=r_1=\gamma$ and $n=1$. This problem was revisited recently in the context of resetting processes and using harmonic traps \cite{santra2021brownian}. In that case, the existence of an optimal switching rate $\gamma$ was shown by numerical simulations. No phase transition occurs in this problem, though. By letting $r_0$ and $r_1$ vary independently, as we have done here, not only lower values of the mean first passage time can be reached in principle, but also a completely different phenomenology is unveiled. This can be summarized by the non-monotonic behavior of the dispersion relation $f(k,r_0)$ with respect to $r_0$, and the fact that this function changes sign when $k$ is larger than a critical value. These two combined features trigger the phase transition to the finite optimal rates $r_0^*$ and $r_1^*$.

We have focused here on the case where the initial state of the potential is "on".
Under this initial condition, we studied how getting the system out of equilibrium by switching the potential off at some small rate $r_1$ compared to the switch-on rate $r_0$ can drastically change the optimal MFPT. Further analysis of other regimes in the parameter space could also be done. When the rate $r_1$ is much larger than $r_0$, for instance, the particle is mostly freely diffusing and our analysis is not valid. In the semi-infinite geometry the MFPT diverge at $r_0=0$ and a perturbation theory is not feasible. The limit of infinite potential stiffness with arbitrary rates $r_0$ and $r_1$  has been previously studied in \cite{Mercado_V_squez_2020}. In this case, our model reduces to the problem of resetting with refractory periods \cite{evans2018effects}, in which after resetting, the particle remains immobile during a mean time $1/r_1$.

If the potential is initially in the "off" state, the full exact solution of the case $n=1$ shows qualitatively different properties whose analysis is more involved \cite{Mercado_V_squez_2020}. There is again a transition at a critical stiffness $k_c'$ (different from $k_c$) above which $r_1^*$ becomes non-zero, and the transition is discontinuous in this case. It would be interesting study this problem for other types of trapping potentials.

The present study could motivate future experiments as well as extensions of the theory to two-dimensional resetting potentials. More complicated potential shapes, not necessarily confining, could be explored as well.
Another interesting problem is the study of the distribution of the work performed by the Brownian particle until it is absorbed. The mean work can then define a search cost, whose optimal parameters might differ from those of the search time.

\vspace{.5cm}
{\bf Acknowledgements:}
GMV thanks CONACYT (Mexico) for a scholarship support and the Laboratoire de Physique Th\'eorique et Mod\`eles Statistiques (LPTMS) for hospitality. We acknowledge support from Ciencia de Frontera 2019 (CONACYT), project \lq\lq Sistemas complejos estoc\'asticos: Agentes m\'oviles, difusi\'on de part\'iculas, y din\'amica de espines\rq\rq \ (Grant 10872).

\appendix

\section{Orstein-Uhlenbeck problem}\label{OU section}

In this Appendix, we consider the problem of a steady harmonic potential. In our model, this case corresponds to setting the potential initially on, $\sigma(t=0)=1$, without turning it off afterwards ($r_1=0$). 

Using the identity $\frac{\partial^2 t_1(x)}{\partial x^2}- k(x-1)\frac{\partial t_1(x)}{\partial x}=e^{\frac{k}{2}(x-1)^2}\frac{\partial}{\partial x}\left(e^{-\frac{k}{2}(x-1)^2}\frac{\partial t_1(x)}{\partial x}\right)$ and setting $r_1=0$ in Eq. (\ref{TP1}) one obtains
\begin{equation}
    e^{\frac{k}{2}(x-1)^2}\frac{\partial}{\partial x}\left(e^{-\frac{k}{2}(x-1)^2}\frac{\partial t_{OU}(x)}{\partial x}\right)=-1,\label{OUMFHT}
\end{equation}
where we have defined $t_{OU}(x)=t_1(x,r_1=0)$ as the rescaled Orstein-Uhlenbeck MFPT (in units of $L^2/D$), corresponding to a Brownian particle in an harmonic potential $v(x)=k(x-1)^2/2$ with an absorbing boundary at $x=0$ and a reflective one at $x=c$ (in units of the length $L$ between the potential minimum and the absorbing wall).

The solution of Eq. (\ref{OUMFHT}) with the boundary conditions  
\begin{eqnarray}
t_{OU}(x=0)&=&0,\\
\frac{\partial t_{OU}(x)}{\partial x}\Big|_{x=c}&=&0,
\end{eqnarray}
is given by
\begin{equation}
\fl    t_{OU}(x)=\sqrt{\frac{\pi}{2 k}}\int_{0}^{x}dy\ e^{\frac{k}{2}(y-1)^2}\left[\mathrm{erf}\left( \sqrt{k/2}(c-1)\right)-\mathrm{erf}\left( \sqrt{k/2}(y-1)\right)\right],
\end{equation}
where $\mathrm{erf}(\cdot)$ is the error function defined as $\mathrm{erf}(x)=\frac{2}{\sqrt{\pi}}\int_{0}^{x} e^{-t^2}dt$.

\section{Numerical solution for the MFHTs \texorpdfstring{$t_0$}{Lg} and \texorpdfstring{$t_1$}{Lg}}\label{finitdif}

We present here a method for obtaining the numerical solution of the system (\ref{TP2})-(\ref{TP1}), which is based on a finite difference scheme for the two-point
boundary value problem. 

We discretize the interval $[0,c]$ into $N$ equal parts, where $N$ is a positive integer. Let $h=c/N$ be the step-size of the grid defined by the points $x_i=ih$, where $0\le i \le N$. In the numerical approximation of $t_0$ and $t_1$, we use the notation  $y_{i}\equiv t_0(x_i)$ and $z_i\equiv t_1(x_i)$, with $0\le i \le N$. 
For the derivatives of $t_0$ at $x$ we use the simple scheme:
\begin{eqnarray}
t_0^{\prime}(x_i)&=&\frac{y_{i+1}-y_{i-1}}{2h}+\mathcal{O}(h^2),\\
t_0^{\prime\prime}(x_i)&=&\frac{y_{i+1}-2y_{i}+y_{i-1}}{h^2}+\mathcal{O}(h^2),
\end{eqnarray}
Similarly for $t_1$,
\begin{eqnarray}
t_1^{\prime}(x_i)&=&\frac{z_{i+1}-z_{i-1}}{2h}+\mathcal{O}(h^2),\\
t_1^{\prime\prime}(x_i)&=&\frac{z_{i+1}-2z_{i}+z_{i-1}}{h^2}+\mathcal{O}(h^2).
\end{eqnarray}
Dropping the terms $\mathcal{O}(h^2)$, the system (\ref{TP1})-(\ref{TP2}) with the harmonic potential $v(x)=k(x-1)^2/2$ reads
 \begin{eqnarray}
     -1&=\frac{y_{i+1}-2y_{i}+y_{i-1}}{h^2}-r_0\left(y_i-z_i\right),\label{TP2discret}\\
      -1&=\frac{z_{i+1}-2z_{i}+z_{i-1}}{h^2}- k(x_i-1)\frac{z_{i+1}-z_{i-1}}{2h}-r_1\left(z_i -y_i\right),\label{TP1discret}
\end{eqnarray}
for $1\le i \le N-1$. Rearranging the terms we get
\begin{eqnarray}
\fl     -h^2&=y_{i+1}-\left(2+h^2r_0\right)y_{i}+y_{i-1}+h^2r_0z_i,\label{TP2discret2}\\
\fl      -h^2&=\left(1-\frac{hk(x_i-1) }{2}\right)z_{i+1}-\left(2+h^2r_1\right)z_{i}+\left(1+\frac{hk(x_i-1) }{2}\right)z_{i-1}+h^2r_1y_i,\label{TP1discret2}
\end{eqnarray}
Imposing the absorbing boundary condition $y_0=0$ and $z_0=0$ at $i=0$, we have, from Eqs. (\ref{TP2discret2})-(\ref{TP1discret2}),
\begin{eqnarray}
     -h^2&=y_{2}-\left(2+h^2r_0\right)y_{1}+h^2r_0z_1,\label{TP2discret3}\\
      -h^2&=\left(1-\frac{hk(x_1-1) }{2}\right)z_{2}-\left(2+h^2r_1\right)z_{1}+h^2r_1y_1.\label{TP1discret3}
\end{eqnarray}
The reflecting boundary conditions at $i=N$ can be enforced by using the first derivative with second-order accuracy
\begin{eqnarray}
t_0^{\prime}(c)&=\frac{3y_{N}-4y_{N-1}+y_{N-2}}{2h}=0,\\
t_1^{\prime}(c)&=\frac{3z_{N}-4z_{N-1}+z_{N-2}}{2h}=0,
\end{eqnarray}
which leads to
\begin{eqnarray}
y_N&=\frac{4y_{N-1}-y_{N-2}}{3},\\
z_N&=\frac{4z_{N-1}-z_{N-2}}{3}.
\end{eqnarray}
Inserting these expressions into Eqs. (\ref{TP2discret2})-(\ref{TP1discret2}) with $i=N-1$, one obtains
\begin{eqnarray}
\fl     -h^2&=-\left(\frac{2}{3}+h^2r_0\right)y_{N-1}+\frac{2}{3}y_{N-2}+h^2r_0z_{N-1},\label{TP2discret4}\\
\fl      -h^2&=-\left(\frac{2}{3}+\frac{2hk(x_{N-1}-1) }{3}+h^2r_1\right)z_{N-1}+\frac{2}{3}\left[1+hk(x_{N-1}-1) \right]z_{N-2}+h^2r_1y_{N-1}.\label{TP1discret4}
\end{eqnarray}
The above relations can be written under the matrix form
\begin{equation}
   \mathbb{A}{\mathbf w}={\mathbf b},\label{FDmatrix}
\end{equation}
where the numerical solution vector is defined as
\begin{equation}
    {\mathbf w}^T=\left(y_1,y_2,\dots,y_{N-1},z_1,z_2,\dots,z_{N-1}\right),
\end{equation}
the constant vector is given by
\begin{equation}
    {\mathbf b}^T=-h^2\left(1,1,\dots,1\right),
\end{equation}
and where the entries of the matrix $\mathbb{A}$ follow from Eqs. (\ref{TP2discret2})-(\ref{TP1discret2}), together with the special cases for $i=1$ and $i=N-1$, given by Eqs. (\ref{TP2discret3})-(\ref{TP1discret3}) and Eqs. (\ref{TP2discret4})-(\ref{TP1discret4}):

\begin{eqnarray}
\fl
\mathbb{A}=\left(
  \begin{array}{ccccccccc}
  -(2+h^2r_0) & 1 & 0 & \cdots & 0 & 0 & h^2r_0 & 0 & 0 \\
   1 & -(2+h^2r_0) & 1 & \cdots & 0 & 0 & 0 & h^2r_0 & 0 \\
  \vdots & \vdots & \vdots & \ddots & \vdots  & \vdots & \vdots & \vdots & \vdots \\
  0 & 0 & 0 & \cdots & \frac{2}{3} & -\left(\frac{2}{3}+h^2r_0\right) &  0 & 0 & 0\\
  h^2r_1 & 0 & 0 & \cdots & 0 & 0  & -(2+h^2r_1) & 1-\frac{hk(x_1-1) }{2} & 0\\
  0 & h^2r_1 & 0 & \cdots & 0 & 0 &  1+\frac{hk(x_1-1) }{2} & -(2+h^2r_1) & 1-\frac{hk(x_1-1) }{2}\\
  \vdots & \vdots & \vdots & \ddots & \vdots  & \vdots & \vdots & \vdots & \vdots \\
  0 & 0 & 0 & \cdots & 0 & h^2r_1 & 0 & 0 & 0 \\
  \end{array}\right.                
\\
  \left.
  \begin{array}{ccc}
   \cdots & 0 & 0 \\
    \cdots & 0 & 0  \\
     \cdots &  \vdots & \vdots  \\
     \cdots & 0 & h^2r_0 \\
  \cdots & 0 & 0 \\
\cdots & 0 & 0 \\
  \cdots &  \vdots & \vdots  \\
 \cdots & \frac{2}{3}\left(1+hk(x_{N-1}-1) \right) & -\left(\frac{2}{3}+\frac{2hk(x_{N-1}-1) }{3}+h^2r_1\right)
  \end{array}\right)
\end{eqnarray}
The solution ${\mathbf w}$ is found directly by numerical inversion of (\ref{FDmatrix}) or
\begin{equation}
    {\mathbf w}=\mathbb{A}^{-1}{\mathbf b}.
\end{equation}
From ${\mathbf w}$ we obtain the numerical solution of the mean first passage times $t_0$ and $t_1$ at any specified starting point $x_i$. It is straightforward to generalize the matrix $\mathbb{A}$ to an arbitrary external potential.

\section*{References}
\bibliographystyle{iopart-num}
\bibliography{Biblio}

\end{document}